\documentclass[11pt,fleqn]{article}
\usepackage{amssymb}

\usepackage{amsmath}

\usepackage{amsmath,amssymb,amsthm}

\begin{document}

\begin{center}

{\Large \textbf{Differential Invariants for Infinite-Dimensional
Algebras}}\footnote{Published in Proceedings of the International
Conference on SPT 2004, Cala Gonone, Sardinia, Italy, May 30 --
June 6,  2004, p. 308--312}

\vskip 18pt {\large \textbf{Irina YEHORCHENKO}}

\vskip 6pt {Institute of Mathematics of NAS Ukraine, 3
Tereshchenkivs'ka Str., 01601 Kyiv-4, Ukraine} E-mail: {\it
iyegorch@imath.kiev.ua}

\end{center}

\begin{abstract}
We present an approach for construction of functional bases of
differential invariants for some infinite-dimensional algebras
with coefficients of generating operators depending on arbitrary
functions. An example for the infinite-dimensional Poincare-type
algebra is given.
\end{abstract}

\section{Introduction}
Our studies in differential invariants (see e.g. \cite{FY})
started from the problem of description of equations invariant
under certain algebras.If we speak about single equations, we can
say that all equations invariant under certain algebras can be
presented as functions of absolute differential invariants.

The theory and methods for searching differential invariants of
finite-dimensional Lie algebras are well-developed. See for the
relevant definitions e.g.\ the classical books
\cite{book1,Ovs-eng}.

All absolute invariants can be presented as functions of
invariants from a functional basis. The number of invariants (of a
certain particular order~$r$) is determined as difference between
the number of all derivatives up to the $r$-th order and both
dependent and independent variables, and of the rank of $r$-th Lie
prolongation of the basis operators of the algebra under
consideration.

The case of infinite-dimensional algebras is more complicated, as
their bases contain infinite (countable) number of operators
(e.g.\ Virasoro and Kac-Moody algebras), or contain infinitesimal
operators having arbitrary functions as coefficients.

However, it appears that despite the name ``infinite-dimensional''
ranks of $r$-th Lie prolongations of basis operators are finite
for each fixed $r$. Unlike finite-dimensional algebras these ranks
do not stabilise, or do not reach any fixed value. Finiteness of
such rank is discussed in \cite{munoz}. Such finiteness is obvious
as the rank of the $r$-th prolongation of the basis operators
cannot exceed the number of all derivatives up to the $r$-th order
and both dependent and independent variables.

Calculation of differential invariants for infinite-dimensional
algebras is specifically interesting in application to equivalence
algebras of classes of differential equations, as knowledge of
such invariants gives criteria for equivalence of different
equations from the same class with respect to local
transformations of variables. In the case of ODE knowledge of such
invariants gives both necessary and sufficient conditions of
equivalence \cite{berth}.

In a number of papers by N.H.~Ibragimov and his coauthors (see
e.g.~\cite{I,IS}) invariants for equivalence algebras of classes
of differential equations are sought for directly.

Here we suggest a systematic procedure that allows considerable
simplification of these calculations. Instead of arbitrary
functions usually found in basis operators of equivalence
algebras, we use expansions of these functions into Taylor series.
We have to remember that we deal with arbitrary functions. Though
they have to be infinitely differentiable (due to commutation
condition in the definition of the Lie algebra), that does not
mean that they are analytical. However, for the purpose of
calculation of differential invariants of infinite-dimensional
algebras we can reasonably limit our consideration by analytical
functions in coefficients of basis operators, and with finite
number of such arbitrary functions in coefficients of basis
operators.

Using expansion of coefficients into series allows replacement of
operators with arbitrary functions with infinite series of
infinitesimal operators without such arbitrary functions. This
approach allows much more straightforward calculation of the
prolongations' rank (in some cases the rank is equal to the number
of variables and derivatives, and then it is easy to see without
any further calculations that there are no absolute invariants of
the respective order).

\noindent
{\bf Statement.} For a fixed order $r$ there is a functional basis
of any Lie algebra, including infinite-dimensional algebras with
finite number of such arbitrary functions in coefficients of basis
operators or with countable infinite sequences of basis operators
with no arbitrary functions.

\section{Differential Invariants for Infinite-Dimensional Poincar\'e-Type
 Algebra}

We will illustrate our approach to searching absolute differential
invariants of infinite-dimensional algebras by the example of
infinite-dimensional Poincar\'e-type algebra that is an invariance
algebra of the eikonal equation.

It is well-known that the simplest first-order relativistic
equation --- the eikonal or Hamilton equation, for  $n$
independent space variables $x_n$ and time variable $x_0$, and
scalar dependent variable $u$,
\begin{equation} \label{eik}
u_\alpha u_\alpha\equiv u_0^2-u_1^2-\cdots-u_n^2=0
\end{equation}
is invariant under the infinite-dimensional algebra generated by
the opera\-tors \cite{FSS}
\begin{equation} \label{eikIA}
X=(b^{\mu\nu}x_\nu+a^\mu)\partial_\mu+\eta(u)\partial_u,
\end{equation}
$-b^{\mu\nu}=b^{\nu\mu}$, $a^\mu$, $\eta$ being arbitrary
differentiable functions on $u$, $\partial_\mu=\partial/\partial
x_\mu$. Usual summation is implied over the repeated Greek
indices: $u_\mu u_\mu=u_0^2-u^2_1-\cdots-u^2_n$. Equation
\eqref{eik} is widely used e.g.\ in geometrical optics.

Here we construct differential invariants of orders 1 and 2.
Instead of the operators \eqref{eikIA}, after expansion of the
functions  $-b^{\mu\nu}=b^{\nu\mu}$, $a^\mu$, $\eta$ into Taylor
series, we can consider the following sequences of operators:
\begin{equation} \label{eikIA1}
J^k_{\mu\nu}=u^k(x_\mu \partial_\nu - x_\nu \partial_\mu), \quad P^k_\mu
= u^k\partial_\mu, \quad P^k_u=u^k\partial_u.
\end{equation}

For the order 1 we have $n+2$ variables and $n+1$ first
derivatives, and the rank of the first prolongation of
\eqref{eikIA} is equal to $2n+3$. There is no absolute invariants,
and one obvious relative invariant $u_\mu u_\mu$. It is
interesting to note that for the first order invariance under
\eqref{eikIA} implies invariance under the dilation operator
$D=x_\mu \partial_\mu$.

The generating set of operators for the first prolongation will be
$$J_{\mu\nu}=x_\mu \partial_\nu - x_\nu \partial_\mu,\ D,\
P_u^0,\ P_\mu^0.$$

Rank of the first prolongation of this set is $2n+3$, and
invariance under prolongations of these operators is equivalent to
invariance under the first prolongation of the
algebra~\eqref{eikIA1}.

Finding such simple generating set (with the rank and number of
operators equal to the rank of the first prolongation) makes
finding invariants much simpler, and would allow using computer
software to do so).

The tensor of the rank 2 \cite{FY}
\begin{equation} \label{i-tensor}
\theta_{\mu\nu}=u_\mu u_{\lambda\nu} u_\lambda+u_\nu
u_{\lambda\mu} u_\lambda-u_\mu u_\nu u_{\lambda\lambda}-u_\lambda
u_\lambda u_{\mu\nu}
\end{equation}
is covariant under the algebra \eqref{eikIA1} (for simplicity of
the definition, we say that a tensor is covariant under a certain
algebra, if all its convolutions are relative or absolute
invariants of this algebra).

Covariance can be checked directly by application of the Lie
algorithm.

Calculation of the rank of the second prolongation of
\eqref{eikIA1} gives that there will be $n$ second-order
invariants:
\begin{equation}
S_k/(u_\mu u_\mu)^{(3/2)k}, \label{inv}
\end{equation}
where $S_k = \theta_{\mu_1 \mu_2}\theta_{\mu_2 \mu_3}\cdots
\theta_{\mu_{k-1} \mu_1}$, $k=1,...,n$.

\smallskip

\noindent {\bf Statement.} The set \eqref{inv} is a functional
basis of second-order absolute differential invariants of the
algebra \eqref{eikIA1}.

\smallskip

Here we can make a comment on sufficiency of consideration of
analy\-ti\-cal functions in \eqref{eikIA} and correctness of the
transition to algebra \eqref{eikIA1}.

We have found a set of functionally independent invariants using
the algebra \eqref{eikIA1} and its rank; the rank of the second
prolongation of \eqref{eikIA} cannot be larger, as otherwise there
would be only a smaller set of functionally independent
invariants.

\section{Conclusion}

Here we present the steps for calculation of functional bases of
absolute differential invariants for infinite-dimensional algebras
with arbitrary functions in basis operators:
\begin{enumerate}
\itemsep=0pt

\item[1.] Expand functions into Taylor series.

\item[2.] Transform the set of operators with arbitrary functions into
discrete infinite set without arbitrary functions.

\item[3.] Find needed prolongations of the algebra.

\item[4.] Calculate rank of the prolongation of the algebra.

\item[5.] Find a minimal ``generating set'' of operators with the rank of
their prolongation equal to that of the prolongation of the
algebra.

\item[6.]  Find a functional basis using the ``generating set''.
\end{enumerate}

Further research in this direction, beside calculation of
invariants for other algebras, includes studying dependence of the
ranks and orders of prolongations, and structures of the
generation sets for the prolongations that may be of interest for
study of the algebraic properties of invariant equations.

\section*{Acknowledgments}

I would like to thank the University of Milan whose grant enabled
me to participate in SPT2004, and Professor G.~Gaeta and SPT2004
Organising Committee for hospitality. I also would like to thank
my colleagues V.~Boyko, R.~Popovych, N.~Ivanova, A.~Zhalij and
A.~Nikitin for fruitful discussions during preparation of this work
and for providing valuable references.

\end{document}